# END-TO-END SPEECH RECOGNITION WITH ADAPTIVE COMPUTATION STEPS

*Mohan Li, Min Liu, Hattori Masanori*

Toshiba (China) R&D Center

## ABSTRACT

In this paper, we present *Adaptive Computation Steps* (ACS) algorithm, which enables end-to-end speech recognition models to dynamically decide how many frames should be processed to predict a linguistic output. The model that applies ACS algorithm follows the encoder-decoder framework, while unlike the attention-based models, it produces alignments independently at the encoder side using the correlation between adjacent frames. Thus, predictions can be made as soon as sufficient acoustic information is received, which makes the model applicable in online cases. Besides, a small change is made to the decoding stage of the encoder-decoder framework, which allows the prediction to exploit bidirectional contexts. We verify the ACS algorithm on a Mandarin speech corpus AIShell-1, and it achieves a 31.2% CER in the online occasion, compared to the 32.4% CER of the attention-based model. To fully demonstrate the advantage of ACS algorithm, offline experiments are conducted, in which our ACS model achieves an 18.7% CER, outperforming the attention-based counterpart with the CER of 22.0%.

## 1. INTRODUCTION

The recent popularity of sequence-to-sequence (seq2seq) models in the field of automatic speech recognition (ASR) has significantly challenged the dominating role of traditional HMM-based ASR systems. Especially, the explorations in end-to-end recurrent neural networks (RNNs) enabled researchers to merge separate components of a traditional system (acoustic, pronunciation and language models) into one integral structure. Such end-to-end models not only reform the way of training by jointly optimizing all of their sub-components without any prior knowledge, but also make it possible for the ASR system to generate outputs at any linguistic level, ranging from phonemes to syllables or even words.

Unlike the HMM-based models predicting a target (HMM-state) for each frame, some seq2seq models recognize utterances by aligning to some parts of the inputs at each decoding step. These models are epitomized by the *attention-based encoder-decoder* networks, upon which groundbreaking progress was made on neural machine translation [1], image caption generation [2] and handwriting synthesis [3] tasks. The architecture was soon introduced to training ASR models, and achieved promising results in phoneme [4, 5] and character [6, 7] recognitions. When exposed to sufficient training data, RNNs equipped with multi-head attention mechanism [8] could even acquire competitive word level performance compared with HMM-DNN hybrids.

In this paper, we propose a novel end-to-end ASR model which dynamically adapts the number of computation steps that are needed to predict a linguistic output. The idea is inspired by the fact that the time we use to recognize a word or a syllable is always less than the time we use to listen to it. Besides, the predictions are only made when we are confident enough on the received acoustic information. Similarly, ASR apparatus takes in speech signals as a sequence of frames, and produces discrete outputs only at certain times. For most of the computation steps, the model may be seen as to ponder and consolidate the information in the input sequence and wait for a proper timing to emit an output with confidence. Therefore, we aim to make the ASR model spontaneously decide the time for making a prediction, and limit the time of decision-making to a small vicinity of input frames to avoid incurring big delays.

As a comparison to the attention-based models, we improve the performance of the end-to-end ASR system in the following aspects:

1. Instead of referring to the entire input sequence, for each output step, our model only focuses on a block of continuous inputs that are acoustically related to the target.
2. The speech is processed in a left-to-right frame-wise manner so that the model is applicable in online occasions.
3. The outputs predicted by the decoder will not only depend on the decoding histories, but on bidirectional contexts.
4. We combine an external RNN language model with the ASR model to make them work in parallel, which keeps the decoding process still in an end-to-end manner.

## 2. RELATION TO PRIOR WORK

Though attention-based models have gained popularity due to the simplicity in training and efficiency in decoding, there are some problems caused by their intrinsic working principles.

First, at each decoding step, the attention mechanism calculates a score for every hidden state (referred to as memory in terminology) of the RNN encoder, which brings huge computation burdens to the system, especially for long utterances. From the acoustic perspective, the information in a feature frame is only correlated to its a few neighbors. Therefore, even if attention is applied to all frames, only a small group of them are making contribution in predicting the target. As proposed in [6, 7], one of the solutions is to slide a window along the memory to restrict the frames to be attended to. However, as the size and stride of the window are empirically estimated from the training data, it is hard to ensure that all target-pronounced inputs are included in the window. This is very likely to happen when the speaking speed varies greatly in a corpus.

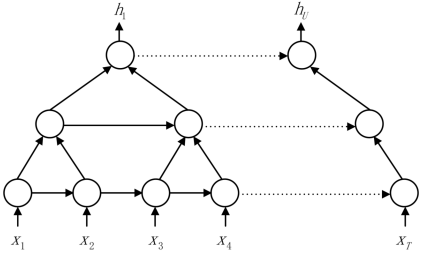

**Fig. 1.** A 3-layer pyramidal RNN encoder. The top two layers totally reduce the frame rate by a factor of 4 ($2^2$).

Moreover, we know that speech recognition is a strictly left-to-right decoding process, so the inputs that have already been attended are not expected to be referred back at subsequent decoding steps. Several attempts have been made to encourage such monotonicity, including penalizing attentions that align to previously observed memory entries [4], predicting the next alignment center given the current attention distribution [9], and replacing the softmax-based attention mechanism with a hard monotonic aligning algorithm [10, 11]. Strategies above effectively force the attention-based decoder to move forward along the utterance, whereas the occurrences of extra complexities make the models even harder to optimize.

The model proposed in this paper will circumvent these problems by processing frames incrementally and predicting targets as soon as sufficient information is gathered from the computation steps. The detailed model architecture will be given in Section 3.

## 3. ADAPTIVE COMPUTATION STEPS FOR SPEECH RECOGNITION

### 3.1 Encoder-Decoder Architecture

The general framework of the model we proposed consists of two recurrent neural networks, referred to as encoder and decoder. Let $\mathbf{x} = (x_1, x_2, ..., x_T)$ denotes the sequence of input feature vectors, and $\mathbf{y} = (y_1, y_2, ..., y_L)$ be the output sequence generated from input $\mathbf{x}$ through the architecture. To begin with, the RNN encoder transforms input $\mathbf{x}$ into a sequence of higher dimensional representations $\mathbf{h} = (h_1, h_2, ..., h_U)$. For standard multi-layer RNNs, $\mathbf{h}$ is in the same length with inputs $\mathbf{x}$, as the network generates a hidden state for each inputting time-step. However, as we know, it usually dozens tens of frames to pronounce a syllable or word, thus keeping the representation for all time-steps might result in redundant information for the rest of model components. To solve this, we use the pyramidal RNN architecture [6, 7, 12, 13] as the encoder, which down-samples the hidden states of the previous layer before feeding them to the next layer. The input to the $i^{th}$ layer at time-step $j$ is calculated as:

$$h_j^i = Recurrency(h_{j-1}^i, [h_{2j-1}^{i-1}, h_{2j}^{i-1}]) \quad (1)$$

Compared to the frame-skipping strategy [14, 15] deployed at the input layer, pyramidal RNN reduces the layer-wise time resolution without losing the continuity of information, which helps to stabilize the following stage of pondering and encourage the model to make quicker decisions. See Figure 1 for an example of such encoders.

The RNN decoder is designed to predict the output sequence $\mathbf{y}$ conditioned on some entries of $\mathbf{h}$, as well as the decoding histories. To be concrete, the probability distribution over $y_i$ is a function of the decoder state $s_{i-1}$, the embedding of the previously decoded $y_{i-1}$, and the context $c_i$ summarized by a *halting layer*:

$$p(y_i) = softmax(Recurrency(s_{i-1}, y_{i-1}, c_i)) \quad (2)$$

The decoder, in a manner, plays the role of the language model as in the traditional ASR system. We achieve this by providing the ground-truth label as the previous prediction $y_{i-1}$ in the training stage. The grammar dependencies are propagated in the decoder hidden states and help to produce rational sentences as possible.

### 3.2 Adaptive Computation Steps

Let $N_i$ be the number of transformations that the encoder performs before predicting a target at decoding step $i$. To determine $N_i$, we first apply a 1-D convolutional neural network (CNN) on the representations $\mathbf{h}$ along encoding time $j$. Thus for each encoder time-step, the CNN extracts an energy vector $e_j$ from a small segment of $\mathbf{h}$ around $h_j$:

$$e_j = \text{Convolution1d}(\tilde{h}_j) \quad (3)$$

where $\tilde{h}_j$ is a window centered at $h_j$. The dimension of $e_j$ depends on the number of output channels of the CNN kernel. Then $e_j$ is projected to a scalar by a sigmoidal halting unit [16], whose activation $a_j$ will be used to determine the *halting probability* $p_j$ of the current computation step:

$$a_j = \sigma(e_j) \quad (4)$$

where $\sigma(\cdot)$ is the logistic sigmoid function. We keep inspecting the accumulation of the $a_j$ along time $j$, and halt the encoder's computation as soon as the sum exceeds 1. Suppose the above described "pondering" process starts from time $j = 1$, and we can calculate $N_i$ as:

$$N_i = \min\left\{n: \sum_{j=1}^{n} a_j \geq 1 - \epsilon\right\} \quad (5)$$

Note that we expect the halting probabilities within the pondering interval to be summed up to 1, thus $a_j$ for the last computation step, named as *remainder*, should be modified to:

$$R_j = 1 - \sum_{j=1}^{N_i-1} a_j \quad (6)$$

Therefore, we summarize the definition of halting probability $p_j$ as:

$$p_j = \begin{cases} R_i & \text{if } j = N_i \\ a_j & \text{otherwise} \end{cases} \quad (7)$$

The $\epsilon$ in equation (5) is a small constant of offset (e.g. 0.01), whose purpose is to allow the encoder to halt computing after one single update, otherwise a minimum of two steps must be taken, as the activation of the sigmoid function can converge to but never reach 1.

After finishing the adapted steps of computation for one decoding time, the resulted halting probabilities turn out to be a distribution over a bunch of continuous representations that make up the target being predicted. All of the entries observed within the computation steps ought to take part in emitting the current output. Thus, a context $c_i$ for blending the information in $h_{1\sim j}$ is given as:

$$c_i = \sum_{j=1}^{N_i} p_j h_j \qquad (8)$$

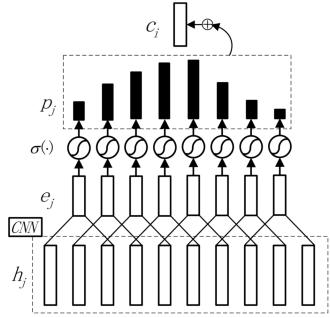

**Fig. 2.** The workflow of ACS algorithm. The halting probability is calculated out of the representations by the halting layer, which consists of a 1-D CNN layer and a sigmoidal unit.

We then feed $c_i$ to the RNN decoder to generate $y_i$ and repeat the procedures above starting from $h_{N_i+1}$ until the end of speech is reached. Figure 2 demonstrates the ACS algorithm for a complete pondering interval.

An alternative approach to determine whether to halt or continue the computation is to draw binary samples from the halting distribution [17], where the decision-making is treated as a Bernoulli process [10, 11]. However, the stochastic gradient estimation precludes the use of backpropagation during model training, and the escalating sampling noise will result in huge derivation for all following decisions [18].

### 3.3 Decoding with Bidirectional Contexts

As shown in equation (2), the prediction of a target depends on the decoding histories and a context vector computed for the current output step. However, we also expect to involve the future context information into the decoding stage to enhance the coherence within the output sequence. Thus, instead of feeding the context vector immediately to the decoder, we keep it until a few following ones are computed by the ACS algorithm. Then we concatenate the context vectors in both directions, and input them to the decoder:

$$P(y_i) = softmax(Recurrency(s_{i-1}, y_{i-1}, [c_{i-w}; \ldots; c_i; \ldots; c_{i+w}])) \qquad (9)$$

where $w$ is the number of neighboring context vectors at each side. We restrict $w$ to a small value (less than 3) to minimize the damage it does to the real-time performance. Note the online bidirectional contexts can only be applied when the encoder and decoder work independently (not the case for attention-based models), so that any number of context vectors can be produced within one decoding step.

### 3.4 Joint Decoding with RNN Language Model

From the perspective of language modelling, the end-to end models suffer from the fact that they are only exposed to the speech transcriptions encountered in the training stage. One of the solutions is to incorporate a language model (LM) that is separately trained on large external corpora. Then the LM can be either embedded in a finite state transducer (FST) to construct a searching network, or used to rescore the beam search results obtained by the end-to-end decoding process [7].

Our proposal is to introduce a RNN-LM, which works in parallel with the model's decoder and produces a LM probability $p_{LM}(y)$ for the output at each decoding step [19]. We interpolate $p_{LM}(y)$ with $p(y|x)$ that is calculated by the decoder to give the final score of the candidate:

$$s(y|x) = \log p(y|x) + \gamma \log p_{LM}(y) \qquad (10)$$

where $\gamma$ is a scaling factor to balance the role of the LM, whose value can be determined on a development set. Here the RNN-LM ingests the grammar of the previously decoded sequence in its cell states, and acts on the subsequent outputs jointly with the decoder, which keeps the decoding process still in an end-to-end manner.

## 4. EXPERIMENTS

### 4.1 Data

All experiments were conducted on an open-source Mandarin speech corpus AIShell-1 [20]. We trained our models on the 150-hour/120098-sentence set with all utterances containing more than 1000 frames removed to save the computation power, and used the 10-hour/14236-sentence development set for early-stopping. The models were finally evaluated on another 7176 sentences presenting speech of approximately 5 hours. The features as the model inputs consist of 1) 24-dimensional energy augmented mel frequency cepstral coefficients (MFCC) with delta and acceleration, 2) 1-dimensional sub-band time average cepstrum coefficient (STAC) with delta, as well as 3) 1-dimensional pitch information with delta and acceleration (total 74 dimensional features). The decoding targets include 7065 Chinese characters and four special tokens as <UNK> (unknown), <PAD> (padding), <SOS> (start of speech) and <EOS> (end of speech).

### 4.2 Training

We started from training an ACS model for online decoding with the unidirectional RNN encoder. Specifically, the encoder was a 3-layer 512-unit pyramidal GRU network with the top two layers totally reducing the frame rate by a factor of 4. The decoder was a 1-layer standard GRU with 256 units. For the halting layer, we used 64 CNN kernels with the width of 3 activated by a rectified linear unit (ReLU). Then, to fully explore the potential of the ACS algorithm, an ACS model with bidirectional RNN encoder was also trained for offline occasions. The encoder used a 3-layer GRU with 256 units for both directions, while the decoder and halting layer shared the same structure with their counterparts in the online ACS model.

All weights were uniformly initialized within the range [-0.1, 0.1]. We trained each model for 30 epochs with Adam optimization algorithm [21] and exponential weight decay strategy. The gradient norm [22] was clipped to 2 for the first 20 epochs and 1 for the rest. Cross-entropy between the predicted Chinese character and its ground-truth label was used to define the loss function.

**Table 1.** Character Error Rate (CER) on HMM-DNN and end-to-end models. The results of attention-based and ACS models were decoded using beam search algorithm with the width of 8.

| Model | CER |
|---|---|
| HMM-Hybrid Models | |
| HMM-DNN [19] | 8.5% |
| Online Character Models | |
| Attention | 34.9% |
| Attention + RNN-LM | 32.4% |
| ACS | 35.2% |
| ACS + Bidirectional Contexts ($w$=1) | 33.5% |
| ACS + Bidirectional Contexts ($w$=1) + RNN-LM | **31.2%** |
| Offline Character Models | |
| Attention | 23.2% |
| Attention + RNN-LM | 22.0% |
| ACS | 21.6% |
| ACS + Bidirectional Contexts ($w$=1) | 19.8% |
| ACS + Bidirectional Contexts ($w$=1) + RNN-LM | **18.7%** |

The baseline in terms of character error rate (CER) came from the s5 recipe of the Kaldi toolkit [23], together with attention-based models for end-to-end level comparison. Note the encoder-decoder architectures in these attention-based models were exactly the same as in our ACS models, for both online and offline cases respectively.

### 4.3 Results

Detailed experiment results are given in Table 1. For the online category, the vanilla ACS model achieves a 35.2 % CER parity with the 34.9% CER of the attention-based model. Additional improvements are obtained by applying bidirectional contexts in the decoding stage, and the resulted CER of 33.5% brings an error reduction rate (ERR) of 3.4% for the ACS model. When introducing an external RNN-LM, the CER further drops to 31.2%, lower than the 32.4% CER achieved by the attention-based model under the same condition.

In offline cases, the advantage of bidirectional RNN encoder makes both ACS and attention-based models outperform their online counterparts. Under such condition, the vanilla ACS model achieving 21.6% CER defeats the attention-based one (23.2%) by an absolute CER reduction of 1.6%, and the disparity rises to 3.3% (18.7% vs. 22.0%) when the bidirectional contexts and RNN-LM are involved. The best result attained by the end-to-end models is still far behind the 8.5% CER of the state-of-the-art HMM-DNN model, but we believe the inferiority will be mitigated by larger training corpora and advanced model architectures.

### 5. DISCUSSIONS

Most of the prevailing attention mechanisms used in speech recognition are referred to as soft attention [4, 5, 6, 7, 8, 9]. It is assumed that the elements required in predicting an output are distributed in multiple inputs. Entries of significant relevance are assigned with higher attention scores to emphasize their roles in generating a transcription. To the contrary, the hard attention mechanism is also adopted by ASR tasks [10, 11], based on the fact that the inputs and outputs hold strictly monotonic alignment properties. Models equipped with hard attention are allowed to focus on single time-steps of the input sequence.

The alignment strategy implied in ACS algorithm combines the advantages of both soft and hard attention mechanisms. On one hand, the aligner calculates a probability of halting for each encoder time-step within the pondering interval, and accordingly summarizes a context vector as does the soft attention-based model. On the other hand, our model constantly inspects the accumulation of the halting probabilities, and makes hard decisions to emit outputs as soon as the sum reaches the threshold. Such eclecticism not only remains the reliabilities by taking advantages of multi-entry alignment used in soft attention mechanism, but also overcomes the problem that models are usually non-differentiable in hard attention architectures.

Moreover, as mentioned in Section 2, the soft attention-based model is computationally expensive as it passes over the entire input sequence at each decoding step, which results in a time complexity of $\mathcal{O}(TU)$ [5], where T and U are the memory and output sequence lengths. In contrast, our model allows for decoding in linear-time complexity $\mathcal{O}(T+U)$ by processing utterances in a monotonically frame-wise manner, where only T terms of halting probabilities are computed throughout the decoding process.

### 6. CONCLUSIONS

In this paper, we propose a novel end-to-end ASR model that adaptively decides the number of computation steps towards the input frames before outputting the corresponding transcription. The model is reported to be more computationally efficient than the attention-based architecture and naturally applicable in online occasions. The results evaluated on an open-source Mandarin test set show that the proposed model achieves better performances than the attention-based models in both online and offline cases.

The core algorithm used to establish the model, known as adaptive computation steps, is realized by a halting layer, which consists of a CNN and a sigmoidal unit. The structure extracts the internal correlations between the adjacent frames, upon which the halting probability is calculated for the model to decide when to make a prediction. We may understand the halting probability as the contribution that an entry makes in predicting a target, which shares the same idea with the attention weights. However, the restriction of its value to (0, 1) imposed by the sigmoid function circumvents the problem caused by normalization, which can only be done after the whole pass of attention is performed over the entire inputs chain. This makes the models with ACS algorithm show great advantages over the attention-based ones in online applications.

Our ACS model performs well in tasks where the output units have relatively clear borders, like Mandarin character (syllable) in this paper. However, for smaller linguistic units that badly overlap in time, such as phoneme, it is difficult for the ACS algorithm to give explicit alignments of the targets all the time. This is also a common shortcoming for the ASR models that decode based on hard alignments. We will leave this problem for future work.

### 7. ACKNOWLEDGEMENTS

All experiments were conducted using Tensorflow libraries [24].
The authors would like to thank Jiaxin Wen for her expertise in linguistics that helped to consolidate the idea proposed in this paper.